\documentclass[aps,pra,showpacs,twocolumn,superscriptaddress]{revtex4-1}
\usepackage{graphicx,amsmath,amssymb}
\usepackage[usenames]{color}
\usepackage{indentfirst}
\usepackage{float}

\usepackage[colorlinks=true, citecolor=blue, urlcolor=blue, linkcolor=blue ]{hyperref}
\hypersetup{breaklinks=true}
\begin{document}
\bibliographystyle{apsrev4-1}
\title{Frequency-renormalized multipolaron expansion for the quantum Rabi model}
\author{Lei Cong}
\affiliation{Center for Interdisciplinary Studies $\&$ Key Laboratory for Magnetism and Magnetic Materials of the MoE, Lanzhou University, Lanzhou 730000, China}
\author{Xi-Mei Sun}
\affiliation{Center for Interdisciplinary Studies $\&$ Key Laboratory for Magnetism and Magnetic Materials of the MoE, Lanzhou University, Lanzhou 730000, China}
\author{Maoxin Liu}
\affiliation{Beijing Computational Science Research Center, Beijing 100084, China}

\author{Zu-Jian Ying}
\email{zjying@csrc.ac.cn}
\affiliation{Beijing Computational Science Research Center, Beijing 100084, China}
\affiliation{CNR-SPIN, I-84084 Fisciano (Salerno), Italy and Dipartimento di Fisica ``E. R. Caianiello", Universit$\grave{a}$  di Salerno, I-84084 Fisciano (Salerno), Italy}

\author{Hong-Gang Luo}
\email{luohg@lzu.edu.cn}
\affiliation{Center for Interdisciplinary Studies $\&$ Key Laboratory for Magnetism and Magnetic Materials of the MoE,
Lanzhou University, Lanzhou 730000, China}
\affiliation{Beijing Computational Science Research Center, Beijing 100084, China}

\begin{abstract}
We present a frequency-renormalized multipolaron expansion method to explore the ground state of quantum Rabi model (QRM). The main idea is to take polaron as starting point to expand the ground state of QRM. The polarons are deformed and displaced oscillator states with variationally determined frequency-renormalization and displacement parameters. This method is an extension of the previously proposed polaron concept and the coherent state expansion used in the literature, which shows high efficiency in describing the physics of the QRM. The proposed method is expected to be useful for solving other more complicated light-matter interaction models.
\end{abstract}

\maketitle

\section{Introduction}\label{intro}
Quantum Rabi model (QRM) describes a two-level system interacting with a single-mode bosonic field \cite{PhysRev.49.324,PhysRev.51.652}. It plays a fundamental role in many fields of physics, such as superconducting circuit quantum electrodynamics (QED) \cite{niemczyk_circuit_2010, wallraff_strong_2004, PhysRevLett.95.060501}, quantum optics \cite{RevModPhys.73.565, scully_quantum_1997}, quantum information \cite{PhysRevA.81.042311, Zhou2016, zhao_engineering_2016}, and quantum computation \cite{PhysRevLett.108.120501},  condensed matter physics \cite{HOLSTEIN1959325}.

Experimentally, the  model has been first realized in the cavity QED systems \cite{RevModPhys.85.623}, in which the coupling strength is quite weak, corresponding to the so-called  weak coupling regime. In this regime, the  rotating-wave approximation (RWA) has been widely employed, which leads to an analytically solvable model, namely, the Jaynes-Cummings (JC) model \cite{1443594}.  The JC model is a basic model in quantum optics which is successful in the understanding of a range of experimental phenomena, such as the well known quantum Rabi oscillation \cite{PhysRevLett.76.1800} and vacuum Rabi mode splitting \cite{PhysRevLett.68.1132}.

Recently, with the advancement of quantum technology \cite{RevModPhys.75.281, englund_controlling_2007}, the so-called strong coupling \cite{wallraff_strong_2004}, ultra-strong coupling \cite{PhysRevLett.105.023601, niemczyk_circuit_2010, PhysRevLett.105.237001, 1367-2630-19-2-023022}  and even the deep strong coupling \cite{PhysRevLett.105.263603} regimes have been experimentally realized in many devices. As a result, the RWA widely used in the literature is no longer valid in these strongly coupling regimes, and thus a full QRM has to be reconsidered in order to describe well the physics observed in these strongly coupling regimes.

It turns out that, despite its simple form, it is not an easy task to fully solve and understand the QRM. Therefore, many approximate methods including adiabatic approximation \cite{PhysRevB.72.195410}, general rotating-wave approximation (GRWA) \cite{PhysRevLett.99.173601} and its extensions \cite{PhysRevA.83.065802, PhysRevA.94.063824, PhysRevA.86.015803}, unitary transformation\cite{PhysRevA.91.053834},
the variational technique \cite{NJP-Meanphoton}, to name just a few, have been proposed. However, it has been shown that each of these approximate methods may be valid in certain limited regime, and an approximate method which is valid in whole parameter regime of the model is still favorable and deserves efforts to develop and improve. In 2011, a remarkable mathematical progress on the integrability of the QRM has been obtained by Braak \cite{PhysRevLett.107.100401} and thus the exact spectra of the QRM have been determined in an analytical way. The exact spectra of the QRM has also been formulated by using Bogoliubov operator technique \cite{PhysRevA.86.023822}. However, in order to explore the full physics of the QRM, it is obviously not enough to only know the spectra of the model, one still needs to know exactly the wavefunction of the model. Therefore, it is still important to explore a simple and straightforward method to study the QRM. On the one hand, this method should be valid in whole parameter regime of the QRM, on the other hand, it should also be convenient to formulate the wavefunction of the QRM. This is our motivation of the present work.

\begin{figure}[tbp]
\begin{center}
\includegraphics[width=0.9\columnwidth]{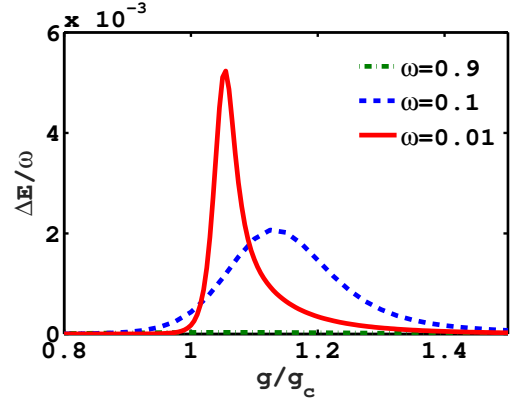}
  \caption{(Color online) Discrepancy of the ground state energy obtained by the polaron and anti-polaron trial wavefunction $E_{gs}$ \cite{PhysRevA.92.053823} in comparison to the exact diagonalization (ED) results $E_{ED}$, which is given by $\Delta E / \omega=(E_{gs}-E_{ED})/\omega$. Here we take $g_{c}=\sqrt{\omega^2+\sqrt{\omega^4+g_{c_0}^4}}$ \cite{PhysRevA.92.053823} and $g_{c_0}=\sqrt{\omega \Omega}/2$.} \label{fig1}
\end{center}
\end{figure}

Very recently, we have introduced a trial wavefunction based on the concept of polaron and anti-polaron picture in order to explore the phase diagram of the QRM \cite{PhysRevA.92.053823}. An important feature of this trial wavefunction is that it provides a unified framework to accurately describe the physics of the QRM both in the weak- and strong coupling regimes. However, in the crossover regime, some errors relatively are still not negligible, particularly, at a low oscillator frequency, as shown in Fig. \ref{fig1}. Therefore, some improvements are still desirable in order to capture more accurate physics in the crossover regime.

How to further improve the calculation accuracy in the crossover regime, particularly, in the low oscillator frequency case? One notes that a variational coherent-state expansion method has been proposed by Bera \emph{et al.} \cite{PhysRevB.89.121108, PhysRevB.90.075110} in the study of the spin-boson model. It was shown that the more the polarons are used, the more accurate the result becomes. Following the idea of the multipolaron expansion, here we propose a variational \emph{frequency-renormalized multipolaron expansion} method to improve the performance of the trial wavefunction based on the polaron and anti-polaron picture. The key difference is that in contrast to Bera \emph{et al.}'s multipolaron expansion, we introduce the frequency renormalization feature. As shown later, the frequency renormalization introduced shows a high efficiency in calculating the energy and wavefunction of the QRM. Therefore, it is expected that our frequency-renormalized multipolaron expansion method is useful in solving more complicated models related to light-matter interaction.

The paper is organized as follows.
In Sec. \ref{model}, the QRM is introduced. In Sec. \ref{method}, we construct our variational method based on the frequency-renormalized multipolaron expansion as the trial ground state wavefunction. In Sec. \ref{results}, we present the results based on the proposed method, and compare them with those obtained by the multipolaron expansion without introducing frequency-renormalized feature. Sec. \ref{conclusion} is devoted to a brief conclusion.

\section{The QRM model}\label{model}
Following the notation in Refs. \cite{PhysRevLett.108.180401, PhysRevB.89.085421, PhysRevA.81.042311, PhysRevA.83.065802, PhysRevA.87.033827, PhysRevA.88.013820}, the Hamiltonian of the QRM model reads ($\hbar = 1$):
\begin{equation}\label{hamiltonian}
H=\frac{\Omega}{2} \sigma_x + \omega \hat{a}^\dagger \hat{a} + g\sigma_{z}{(\hat{a}^\dagger + \hat{a})},
\end{equation}
where $\Omega $ is the qubit energy level splitting,  $\sigma_{x,z}$ is the Pauli matrix to describe the qubit, $\hat{a}^{\dagger}$ and $\hat{a}$ are the bosonic creation and annihilation operators, respectively, of the bosonic mode with frequency $\omega$, and $g$ denotes the coupling strength between the qubit and the bosonic mode.

In terms of the quantum harmonic oscillator with dimensionless formalism $\hat{a}^\dag=(\hat{x}-i\hat{p})/\sqrt{2}$, $\hat{a}=(\hat{x}+i\hat{p})/\sqrt{2}$, where $\hat{x}=x$ and
$\hat{p}=-i \frac{\partial}{\partial x}$ are the position and momentum operators respectively, the model can be rewritten as \cite{PhysRevB.89.085421}
\begin{equation}\label{eq2}
H =\sum_{S_z=\pm} (h^{S_z}|S_z\rangle \langle S_z|+\frac{\Omega}{2}|S_z\rangle \langle\bar{S_z}|)+\varepsilon_0,
\end{equation}
where, $S_z=\pm_z$ represents the eigenvalue of $|\pm_z\rangle$ in z-direction, $\bar{S_z}=-{S_z}$, $h^{\pm }= {\omega}(\hat{p}^2  + (\hat{x}\pm g')^2)/2$, $\varepsilon_0=-\omega({g'}^2+1)/2$ is a constant, and $g'=\sqrt{2} g/\omega$. For simplicity, we take $\Omega=1$ as the units of energy here and hereafter.

\section{Frequency-renormalized multipolaron expansion method}\label{method}
Consider the parity operator $\Pi=\sigma_x e^{a^{\dagger}a}$, one has $[H, \Pi]=0$. Such a $\mathbb{Z}_2$ symmetry leads to a decomposition of the state space into just two subspaces with odd and even parity, respectively. Since the ground state has an odd parity, one takes the trial wavefunction of the QRM in the position representation as
\begin{equation}
|G\rangle= \frac{1}{\sqrt{2}}  ( \Psi^{+}(x) |+_z\rangle - \Psi^{-}(x)|-_z\rangle ),
\end{equation}
where $\Psi^{\pm}$ is the wavefunction associated with the spin state $|\pm_z\rangle$, and we have
\begin{equation} \label{gwf}
\Psi^{+}(x) =\Psi^{-}(-x).
\end{equation}

Based on the polaron picture \cite{PhysRevA.92.053823, PhysRevB.89.121108, PhysRevB.90.075110}, they can be expanded as
\begin{equation} \label{exp}
\Psi^{\pm}(x) =\sum_{n=1}^N C_n \varphi_n^{\pm}(x),
\end{equation}
where $C_n$ is the coefficient and $N$ is the number of polaron used. Here $\varphi_n^{\pm}(x) = \phi_0(\xi_{n}\omega, x {\pm} \zeta_{n}g')$ denotes the $n$th polaron which is given by deformed oscillator ground state wavefunction $\phi_0(\omega,x)$ with the frequency renormalization parameter $\xi_n$ and shifted position parameter $\zeta_{n}$.

Thus, Eq. (\ref{gwf}) can be rewritten as
\begin{eqnarray} \label{trial}
&& |G\rangle = \frac{1}{\sqrt{2}} \sum_{n=1}^N C_n \left(\phi_0(\xi_{n}\omega,x+\zeta_{n}g') |+_z\rangle \right.\nonumber\\
&&\hspace{2cm} \left. - \phi_0(\xi_{n}\omega,x-\zeta_{n}g')|-_z\rangle \right),
\end{eqnarray}
which is the starting point of the present work. Due to the deformed polaron introduced, we call Eq. (\ref{trial}) as frequency-renormalized multi-polaron expansion (FR-MPE). Actually, if one takes $N=2$, Eq. (\ref{trial}) recovers the previous polaron and anti-polaron wavefunction which has been used to explore the ground state phase diagram of the Rabi model \cite{PhysRevA.92.053823}. On the other hand, if one takes $\xi_n=1$, i.e., the frequency-renormalization factor is not considered, Eq.\eqref{trial} is the single mode version of the coherent state expansion used in the study of the spin-boson model \cite{PhysRevB.89.121108, PhysRevB.90.075110}, since a coherent state is a displaced oscillator state in the $\hat x$ representation.

The variational parameters introduced in Eq.\eqref{trial}, i.e., $C_n$, $\xi_{n}$ and $\zeta_{n}$ ($n = 1, \cdots,N$), can be determined by minimizing the ground state energy $E_{G}=\langle G|H|G\rangle$ (see Appendix \ref{app1} for a detailed derivation), subject to the constraint of wavefunction normalization $\langle G|G\rangle=1$, under which the number of variational parameters will be $3 \times N-1$. In order to determine the variational parameters, we first adopt simulated annealing algorithm \cite{Kirkpatrick1984, Hwang1988} to search the rough values of the variational parameters. Then we use pattern search algorithm \cite{Hooke:1961:DSS:321062.321069, davidon1991} to refine these variational parameter values in order to further minimize the ground state energy. The combination of these two algorithms is found to be sufficient to determine the variational parameters with high efficiency and high precision.

\section{numerical results and discussion}\label{results}
\subsection{The result with $N=4$}
\begin{figure}[tbp]
  \includegraphics[width=0.9\columnwidth]{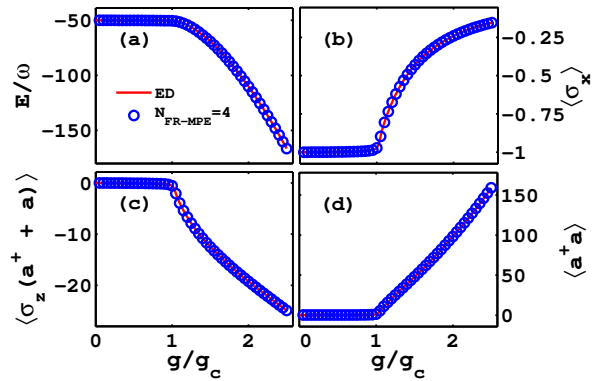}
  \caption{(Color online) Ground-state physical quantities as functions of the coupling strength $g/g_c$. (a) The ground state energy. (b) The spin polarization $\langle \sigma_x\rangle$. (c) The correlation function $\langle \sigma_z(a^+ + a)\rangle$. (d) The mean photon number $\langle a^+a \rangle$. The red lines are the exact diagonalization results taken as benchmarks, and the blue circles are our results obtained by our frequency-renormalized multipolaron expansion given by Eq. (\ref{trial}). Here we take $\omega=0.01$ and $N = 4$. }\label{fig2}
\end{figure}

First of all, we present some results to show the high precision of our method. As shown in Fig. \ref{fig1}, the lower the oscillator frequency is, the larger the error near $g_c$ is. Moreover, the main error is located around $g_c$. Therefore, here we consider the case of $\omega = 0.01$ and take $N=4$. Meanwhile, we compare the obtained results with those with numerical exact diagonalization. In Fig. \ref{fig2} we show various physical quantities including (a) the ground state energy, (b) the spin polarization $\langle \sigma_x\rangle$, (c) the correlation $\langle \sigma_z (a^{\dagger} + a) \rangle$ and (d) the mean photon number $\langle a^{\dagger}a \rangle$ as a function of the coupling strength. Due to the afore-mentioned reason, hereafter we limit ourself to the region around $g_c$. It is found that the agreement is quite good, which will be further discussed in the next subsection. This result confirms the high precision of our method, even in the low oscillator frequency regime. In addition, our method also exhibits high efficiency since only two pairs of polaron and anti-polaron ($N=4$) have been considered here, which indicates that the deformed polaron picture \cite{PhysRevA.92.053823} is a good starting point to capture the physics of the QRM.

\subsection{The high efficiency of the frequency-renormalized multipolaron expansion}
\begin{figure}
  \includegraphics[width=0.9\columnwidth]{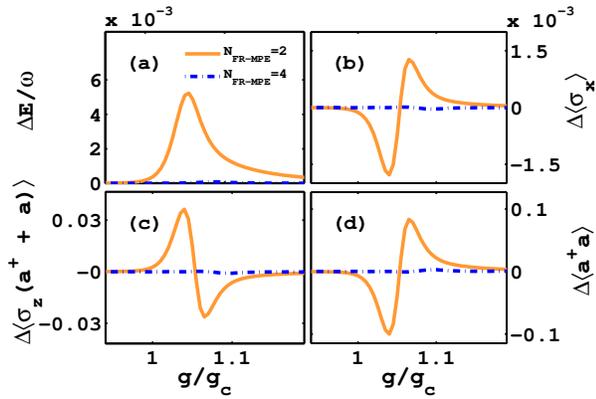}
  \caption{(Color online) Comparison of errors of various physical quantities between $N=4$ and $N=2$. The yellow lines are the result of $N=2$ and the dash dot lines represent the result of $N=4$. (a) Error in the ground state energy with respect to the ED result $\Delta E / \omega=(E-E_{ED})/\omega$. (b) Error in the tunneling strength $\Delta \langle \sigma_x\rangle={\langle \sigma_x\rangle}-{\langle \sigma_x\rangle}_{ED}$. (c) Error in the correlation function $\Delta \langle \sigma_z(a^+ + a)\rangle={\langle \sigma_z(a^+ + a)\rangle}-{\langle \sigma_z(a^+ + a)\rangle}_{ED}$. (d) Error in the mean photon number $\Delta \langle a^+a \rangle={\langle a^+a \rangle}-{\langle a^+a \rangle}_{ED}$. Here we take $\omega=0.01$.} \label{fig3}
\end{figure}

In order to confirm the high efficiency of the frequency-renormalized multipolaron expansion method, we compare the result with $N=4$ with that of $N=2$ with $\omega = 0.01$ in Fig. \ref{fig3}. As also mentioned above, the case of $N=2$ recovers the previous polaron and anti-polaron picture \cite{PhysRevA.92.053823}. From Fig. \ref{fig3} it is noted that the results for $N=4$ have a vanishing small error in comparison to that of $N=2$. Technically, we just increase an additional pair of polaron and anti-polaron beyond that of $N=2$, leading to a dramatic improvement, which shows the high efficiency of the polaron and anti-polaron basis. Physically, it is not difficult to understand why the polaron and anti-polaron basis is so high efficient. This is because that the polaron and anti-polaron as a basic ingredient in describing the ground state wavefunction of the QRM is able to capture the essential physics of the model. The anti-polaron component originates naturally from the tunneling feature \cite{PhysRevB.89.085421, PhysRevA.92.053823} of the QRM. By the same reason, the additional pair of polaron and anti-polaron also originate from the high-order effect of the tunneling feature, which reflects the many-body effect in the QRM. This point is more clear in the discussion of the ground state wavefunction below.

\begin{figure}
  \includegraphics[width=0.9\columnwidth]{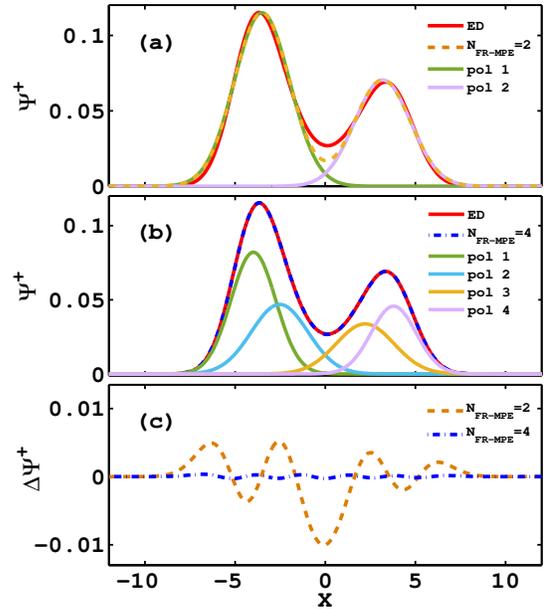}
  \caption{(Color online) Comparison of ground state wave function $\Psi^{+}$ for $|+ \rangle$-component obtained by the frequency-renormalized multipolaron expansion with $N=2$ and $N=4$. The numerical exact result is taken as the benchmark. (a) $N_{FR-MPE}=2$. (b) $N_{FR-MPE}=4$. (c) Error of $\Psi^{+}$ in comparison to the numerical exact one. $\Delta \Psi^{+}=\Psi^{+}_{FR-MPE}-\Psi^{+}_{ED}$. The other parameters are $g/g_c=1.05$ and $\omega=0.01$.}\label{fig4}
\end{figure}

Figure \ref{fig4} shows the ground state wavefunction of the QRM for $N=4$ and its polaron components. As a comparison, we also present the result for $N=2$. For simplicity, we only provide the $\Psi^+$-component, and the $\Psi^-$ has a similar behavior due to the odd parity. From Fig. \ref{fig4}(a), although the case of $N=2$ captures the nature of the wavepacket separation for $g/g_c = 1.05$, the error from the numerical exact result is still obvious, as shown in Fig. \ref{fig4}(c) as yellow dashed line. In particular, in the region around $x=0$, the error is more obvious. This is because that after the wavepacket becomes separated, a pair of polaron and anti-polaron is not sufficient to describe the region away from the positions of the polaron and antipolaron, for example, the region of $x=0$. Therefore, according to the idea of the frequency-renormalized multipolaron expansion, one increases an additional pair of polaron and anti-polaron, not only the region away from the positions of the polaron and anti-polaron, the whole wavefunction has a vanishing small error in comparison with the exact one, as shown in Fig. \ref{fig4}(c) as blue dash-dotted line. This result indicates that the frequency-renormalized multipolaron expansion is highly efficient in calculating the wavefunction of the QRM, as a result, is able to calculate accurately the physical observables, as shown in Fig. \ref{fig3}.

\subsection{The importance of the frequency renormalization}
Based on the scheme of multipolaron expansion introduced by Bera \emph{et al.} in their coherent state expansion (CSE) method \cite{PhysRevB.89.121108, PhysRevB.90.075110}, one of the important features of our method is to introduce the frequency renormalization feature. The physical background of the introduction of the frequency renormalization is the change of the effective potentials induced by the tunneling between these two energy levels, which is the origin of the anti-polaron. The existence of the effective potentials would, of course, modify the frequency of the oscillator states in the QRM. In order to show the importance of the frequency renormalization, in the following we compare our method with the frequency renormalization with the simple multipolaron expansion without the frequency renormalization. The latter is called the CSE method below, which has been used to study the spin-boson model. Here we employ it to the QRM.

\begin{figure}
  \includegraphics[width=0.9\columnwidth]{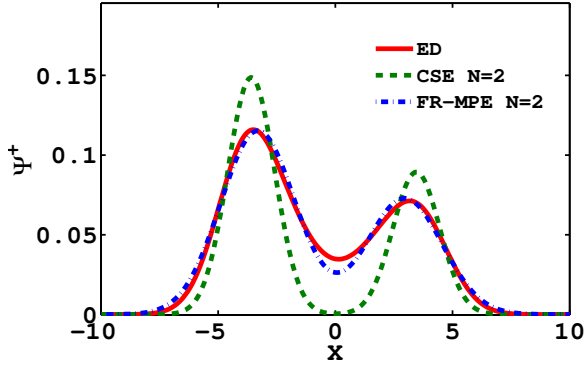}
  \caption{(Color online) Comparison of the ground state wavefunctions obtained by the multipolaron expansion with and without frequency renormalization with the numerical exact result for $N=2$. The other parameters are $g/g_c=1.05$ and $\omega=0.01$.}\label{fig5}
\end{figure}

\begin{figure}
  \includegraphics[width=0.95\columnwidth]{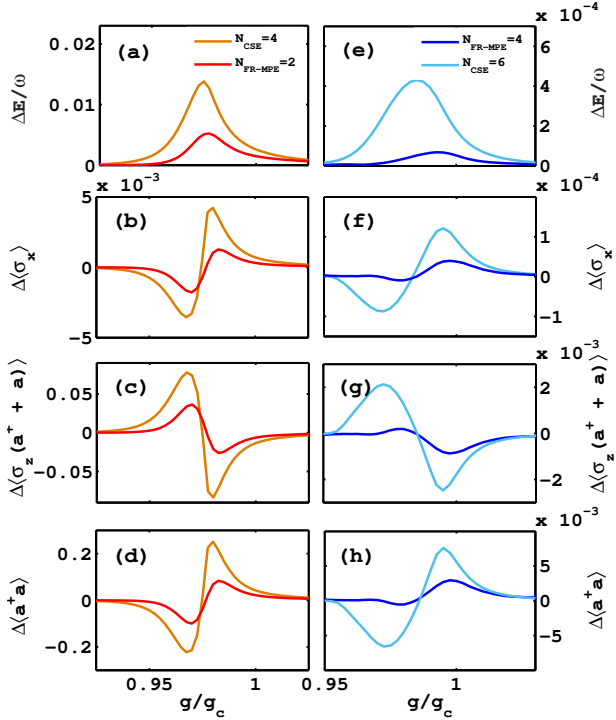}
  \caption{(Color online) Comparison of the ground state energy and the other physical observables obtained by the multipolaron expansion method with and without frequency renormalization. Left column: comparison between $N_{FR-MPE}=2$ and $N_{CSE}=4$. Right column: comparison between $N_{FR-MPE}=4$ and $N_{CSE}=6$. For the left column, the variational parameters for our scheme is 5 and that of the CSE used is 7. In the right column the variational parameters used are the same, both are 11. The errors are defined  the same as those in Fig. \ref{fig3}.}\label{fig6}
\end{figure}

Firstly, we consider the ground state wavefunction for $\Psi^+$-component obtained by these two methods, as shown in Fig. \ref{fig5}, and also present the numerical exact wavefunction for comparison. It is quite obvious that the result with the frequency renormalization agrees much better with the exact one than that without the frequency renormalization. The comparison indicates that the frequency renormalization is indeed an important feature in describing the ground state physics of the QRM. This feature is even more important than simply increasing the number of the polaron and anti-polaron. This can be clearly seen from Fig. \ref{fig6}, showing the comparison of the ground state energy and the other physical observables between the results with frequency renormalization for $N_{FR-MPE}=2,4$ and without frequency renormalization for $N_{CSE}=4,6$, respectively. Obviously, our method has much higher efficiency than that without the frequency renormalization. Moreover, the corresponding variational parameters in our scheme are less than or equal to those of the CSE.

\section{Conclusions} \label{conclusion}
Based on the polaron and anti-polaron picture, we proposed a frequency renormalization multipolaron expansion method to improve significantly the ground-state wavefunction of QRM, as a result, the ground-state energy and other physical quantities have also been significantly improved, in particular, near the crossover region. In comparison to the coherent state expansion given by Bera \emph{et al.} \cite{PhysRevB.89.121108, PhysRevB.90.075110}, our method shows higher efficiency, in which the frequency renormalization plays an important role. Physical origin of the frequency renormalization is due to the deformed potential induced by the tunneling between two energy levels. This method can be applied to other more complicated quantum models like the spin-boson model \cite{PhysRevB.89.121108, PhysRevB.90.075110}, multi-qubit QRM \cite{PhysRevA.81.012105,PhysRevA.93.052305}, anisotropic QRM \cite{PhysRevX.4.021046}, and  biased or asymmetric Rabi model \cite{1751-8121-50-8-084003} and so on. The further works are in progress.

\section*{Acknowledgments}
We acknowledges useful discussion with Gao-Yang Li, Fu-Zhou Chen, Chong Chen and Chen Cheng. This work was supported by NSFC (Grants No. 11325417, No. 11674139) and PCSIRT (Grant No. IRT-16R35). ZJY also acknowledges the financial support of the Future and Emerging Technologies (FET) Programme within the Seventh Framework Programme for Research of the European Commission, under FET-Open Grant No.: 618083 (CNTQC).

\appendix
\section{The ground state energy of FR-MPE}\label{app1}
In this appendix, we present the main steps to get the ground state energy. 
The ground state $|G\rangle$ of FR-MPE is 
\begin{equation}
\begin{split}
|G\rangle = \frac{1}{\sqrt{2}} \sum_{n=1}^N C_n \big{(} \varphi^{+}_n |+_z\rangle - \varphi^{-}_n|-_z\rangle \big{)}.
\end{split}
\end{equation}

The ground state energy of FR-MPE is given by
\begin{equation}\label{A3}
\begin{split}
E_{FR-MPE} & =\langle G|H|G\rangle\\
&=\frac{1}{2}\sum_{n,m}^N C_{n}C_{m} \bigg[ \big( \langle \varphi^{+}_n | h^{+} |  \varphi_{m}^+ \rangle +\langle \varphi^{-}_n | h^{-} |  \varphi^{-}_m \rangle \big)\\
&-\frac{\Omega}{2}\big(\langle{\varphi^{+}_n}|{\varphi^{-}_m}\rangle+\langle{\varphi^{-}_n}|{\varphi_{m}^+}\rangle\big)\bigg]-\frac{1}{2}\omega({g^{'}}^2+1)\\
&=\sum_{n,m}^N  C_{n}C_{m} \langle \varphi^{+}_n |h^{+}|  \varphi_{m}^+ \rangle
-\frac{\Omega}{2}\sum_{n,m}^N  \langle \varphi^{+}_n | \varphi^{-}_m \rangle\\
&-\frac{1}{2}\omega({g^{'}}^2+1).\\
\end{split}
\end{equation}

\subsubsection{Calculation of each term}
For the first term in Eq. (\ref{A3})
\begin{equation}\label{A4}
\begin{split}
h_{n m}^{+}&= \langle \varphi^{+}_n| h^{+} |  \varphi^{+}_m \rangle\\
&=\frac{1}{2} \omega \langle \varphi_{n}^{+} |\big(\hat{p}^2+(\hat{x}+g^{'})^2\big) |\varphi_{m}^{+} \rangle\\
&=\frac{1}{2} \omega \langle \varphi_{n}^{+} | (-\frac{\partial^2{}} {\partial{x^2}}+{x}^2+2{x}{g^{'}}+{g^{'}}^2) | \varphi_{m}^{+} \rangle .\\
\end{split}
\end{equation}
For simplicity we have assumed the unit $\hbar = m = 1$.

We first give the first term in Eq. (\ref{A4}).
\begin{equation}
\begin{split}
\langle \varphi_{n}^{+} |(-\frac{\partial^2{}}  {\partial{x^2}}) | \varphi_{m}^{+} \rangle =&-\langle \varphi_{n}^{+} | (4D^2x^2+4DEx+E^2\\
&+2D) | \varphi_{m}^{+} \rangle,\\
\end{split}
\end{equation}
where, the coefficients $D$ and $E$ are introduced for the simplicity of the formulation, they are defined as
\begin{equation}
\begin{aligned}
D&=-\frac{1}{2}\xi_{m};  E&=-g^{'}\xi_{m}\zeta_{m}.\\
\end{aligned}
\end{equation}
So, the expression of Eq. (\ref{A4}) is
\begin{equation}
\begin{aligned}
h_{n m}^{+}&=\frac{1}{2}\omega \langle \varphi_{n}^{+} | \big[(1-4D^2)x^2+\\
&(2g^{'}-4DE)x-E^2-2D+{g^{'}}^2) |  \varphi_{m}^{+} \rangle ,\\
\end{aligned}
\end{equation}

in which,
\begin{equation}
\begin{aligned}
S_{n m}&=\langle{\varphi_{n}^{+}}(x)|{\varphi_{m}^{+}}(x)\rangle\\
&=\sqrt{2}\big[ \frac{\xi_{n}\xi_{m}}{(\xi_{n}+\xi_{m})^2}\big]^{\frac{1}{4}}%
e^{\big( -\frac{ (\zeta_{n}-\zeta_{m})^2{g^{'}}^2\xi_{n}\xi_{m} }  {2(\xi_{n}+\xi_{m})}\big)},\\
\end{aligned}
\end{equation}
\begin{equation}
\begin{aligned}
{\langle{\hat{x}}\rangle}_{n m}&=\langle{\varphi_{n}^{+}}(x)|\hat{x}|{\varphi_{m}^{+}}(x)\rangle\\
&= S_{n m}\frac{-\xi_{m}\zeta_{m}-\xi_{n}\zeta_{n}}{\xi_{n}+\xi_{m}}  g^{'},\\
\end{aligned}
\end{equation}
and
\begin{equation}
\begin{aligned}
{\langle{\hat{x}}^2\rangle}_{n m}&=\langle \varphi_{n}^{+} | {\hat{x}}^2 | \varphi_{m}^{+} \rangle \\
&=S_{n m} [\frac{1}{\xi_{n}+\xi_{m}} + \big(\frac{{(\xi_{m}\zeta_{m}+\xi_{n}\zeta_{n}){g^{'}}}}{\xi_{n}+\xi_{m}}\big)^2].\\
\end{aligned}
\end{equation}

Until now, we can get the first term in Eq. (\ref{A3}) completely.

For the second term in Eq. (\ref{A3})
\begin{equation}
\begin{aligned}
S_{n\overline m}&=\langle{\psi^{+}_n}|{\psi^{-}_m}\rangle=\langle{\varphi_{n}^{+}}(x)|{\varphi_{m}^{+}}(-x)\rangle\\
&=\sqrt{2}\big[ \frac{\xi_{n}\xi_{m}}{(\xi_{n}+\xi_{m})^2}\big]^{\frac{1}{4}}e^{\big( -\frac{ (\zeta_{n}+\zeta_{m})^2{g^{'}}^2\xi_{n}\xi_{m}}{2(\xi_{n}+\xi_{m})}\big)},\\
\end{aligned}
\end{equation}

So we finally can get the ground state energy.
\subsubsection{Normalization condition}
Besides of the above formulation, we still have the normalization condition which describes the relationships between the parameters.
\begin{equation}
\begin{aligned}
\langle G|H|G\rangle&=\frac{1}{{2}}\sum_{n,m}^N C_{n}C_{m} \big(\langle{\psi^{+}_n}|{\psi^{+}_m}\rangle+\langle{\psi^{-}_n}|{\psi^{-}_m}\rangle\big)\\
&=\sum_{n,m}^N C_{n}C_{m} \big(\langle{\psi^{+}_n}|{\psi^{+}_m}\rangle\big)\\
&=1.
\end{aligned}
\end{equation}

The derivation of the ground state energy of CSE and FR-MPE are nearly the same. Since in the $\hat x$ representation, a coherent state is a displaced ground state of the oscillator. We can get the CSE result by simply setting frequency renormalization factor $\xi=1$ in FR-MPE. So here we present the derivation of FR-MPE result only.

%

\end{document}